# Quantum Capacitance Induced Non-local Electrostatic Gating Effect in Graphene

Aolin Deng[1,2], Cheng Hu[1,2], Peiyue Shen[1,2], Xingdong Luo[1,2], Jiajun Chen[1,2], Bosai Lyu[1,2], Kenji Watanabe[3], Takashi Taniguchi[4], Qi Liang[1,2], Jie Ma[5*], Zhiwen Shi[1,2*]

[1]Key Laboratory of Artificial Structures and Quantum Control (Ministry of Education), Shenyang National Laboratory for Materials Science, School of Physics and Astronomy, Shanghai Jiao Tong University, Shanghai, China.

[2]Collaborative Innovation Center of Advanced Microstructures, Nanjing, China.

[3]Research Center for Functional Materials, National Institute for Materials Science, 1-1 Namiki, Tsukuba 305-0044, Japan.

[4]International Center for Materials Nanoarchitectonics, National Institute for Materials Science, 1-1 Namiki, Tsukuba 305-0044, Japan.

[5]Key Lab of Advanced Optoelectronic Quantum Architecture and Measurement (MOE), School of Physics, Beijing Institute of Technology, Beijing 100081, China.

*To whom correspondence should be addressed. Email: zwshi@sjtu.edu.cn, majie@bit.edu.cn

**Abstract**

Electrostatic gating lies in the heart of modern FET-based integrated circuits. Usually, the gate electrode has to be placed very close to the conduction channel, typically a few nanometers, in order to achieve efficient tunability. However, remote control of a FET device through a gate electrode placed far away is always highly desired, because it not only reduces the complexity of device fabrication, but also enables designing novel devices with new functionalities. Here we report experimental observations of a non-local gating effect in graphene using both near-field optical nano-imaging and electrical transport measurement. We found that with assistance of absorbed water molecules, the charge density of graphene can be efficiently tuned by a local-gate placed over 30 μm away. The observed non-local gating effect is initially driven by an in-plane electric field established between graphene regions with different charge densities due to the quantum capacitance near the Dirac point in graphene. The nonlocality is further amplified and largely enhanced by absorbed water molecules through screening the in-plane electric field and expending the transition length. Our studies reveal novel non-local phenomenon of Dirac electrons, and pave the way for designing electronic devices with remote-control using 2D materials with small density of states.

# 1. Introduction

Similar to the long-range gravitational interaction that governs the universe, the Coulomb interaction in vacuum is also long-range and non-local. However, in condensed matter systems, the long-range Coulomb potential $\sim \frac{1}{r}$ is usually largely screened, and transformed to the short-range Yukawa potential $\sim \frac{e^{-r/r_0}}{r}$, where $r_0$ is the screening length[1]. For example, in copper the strong screening from free carriers leads to a very short screening length of ~0.55 Å. In contrast, the screening can be rather weak in two-dimensional (2D) graphene due to two reasons. Firstly, the reduced dimensionality from 3D to 2D restricts the screening only taking place in a 2D plane. Secondly, the vanished density of states (DOS) near the Dirac point lowers the amount of free carriers and further reduces the screening effect. Consequently, an in-plane electric field can exist inside and around the graphene plane, and the screening length even diverges at Dirac point[2-3]. As a result, long-range interactions and non-local phenomena start to emerge. Several intriguing non-local effects have been observed in graphene and other low-dimensional materials, such as non-local effects in graphene plasmons[4-5] and long-range field effect in WTe$_2$ device[6], etc. The diminished DOS near the Dirac point in graphene also leads to observable quantum capacitance effects[7-8], because the quantum capacitance of a material is directly proportional to its DOS at the Fermi surface[9]. As a result, if charged with electrons or holes, the graphene chemical potential $\mu_c$ (or Fermi level) will vary accordingly, with a speed inversely proportional to the DOS. On the other hand, the electrochemical potential ($\mu_{ec}$), which is the sum of electrical potential ($\mu_e$) and chemical potential ($\mu_c$), $\mu_{ec} = \mu_e + \mu_c$, has to maintain a constant across the whole piece of graphene in equilibrium[10]. If one manages to dope partial region of a whole sheet of graphene, the electrical potential $\mu_e$ of the doped region will be different from that of the undoped region, and a prominent in-plane electric field will emerge between the doped and undoped regions. Such a quantum capacitance induced in-plane electric field together with the decreased screening in graphene may lead to very unusual physical phenomena.

Here, we report experimental observations of a non-local electrostatic gating effect in graphene using both scanning near-field optical microscopy (SNOM)[11-13] and electrical transport measurements[14-15]. Remarkably, the charge density of graphene region 30 μm away from the local gate can be efficiently tuned. The observed non-local gating effect is essentially driven by the quantum capacitance of graphene, and further largely enhanced by absorbed polarized water molecules on graphene surface. Numerical simulations and calculations reproduce the observed non-local gating phenomenon. The reported non-local gating effect provides a deep understanding of in-plane screening from Dirac electrons as well as exciting possibilities for fabricating novel electronic devices with remote gate control.

## 2. Results

**2.1 Observation of non-local electrostatic gating with near-field infrared nanoscopy.**

To investigate the novel non-local gating effect in graphene, graphene device with a local gate was designed and fabricated on Si substrate with a 285-nm-thick $SiO_2$ layer[16-17], as shown in **Figure 1**a,b. The device consists of a large piece of top-layer graphene, a smaller piece of bottom-layer graphene and a nanometer-thick dielectric hexagonal boron nitride (hBN) layer in between, forming a parallel-plate capacitor. Since the bottom-layer graphene overlaps with only part of the large top-layer graphene, it is named a local gate and expected to be able to tune the charge density of the overlapped part. A local-gate voltage ($V_{LG}$) is applied to the bottom-layer graphene, while the top-layer graphene is electrically grounded. Note that the silicon substrate is also grounded to eliminate its gating effect on the top-layer graphene.

The spatial distribution of charge carriers of the top-layer graphene was measured by a home-made scanning near-field optical microscope (SNOM) as shown in Figure 1b. A $CO_2$ laser beam with 10.6 μm wavelength was focused onto a gold-coated atomic force microscope (AFM) tip. The enhanced light field with large momenta at the tip apex can excite graphene plasmons[12-13, 18-20]. The plasmon wavelength can be obtained by measuring the period of the plasmon interference pattern near graphene edges. For a fixed excitation frequency, there is a one-to-one correspondence between the plasmon wavelength $\lambda_p$ and the charge density $n$ as[12-13, 21] $\lambda_p = \frac{8\pi e^2 v_F \sqrt{\pi n}}{\hbar \omega^2 [1+\varepsilon_{sub}(\omega)]}$. Therefore, the distribution of charge density of the graphene flake can be extracted from the near-field infrared nano-images of plasmons, with a typical spatial resolution of ~20 nm determined by the tip size of SNOM.

Figure 1c shows the near-field nano-images under different local-gate voltages $V_{LG}$ ranging from 0 V to -5 V. The bright bottom-left region presents the overlapped area, while the bottom-right and top-left show the non-overlapped top- and bottom-layer graphene, respectively. With increasing the local-gate voltage in the negative side (while the top layer graphene and silicon substrate keep grounded), the top-layer graphene plasmon wavelength $\lambda_p$ increases evidently in both overlapped and non-overlapped areas. We extract the plasmon wavelength $\lambda_p$ and charge density *n* at a specific position (denoted by the blue dashed line) of the non-overlapped area under different gate voltages, and plot them in Figure 1d. Clearly, the plasmon wavelength in the non-overlapped region is tuned efficiently by the remote local gate, reflecting a systematic change of charge density. Note that the change of charge density in the non-overlapped graphene area is very unusual, considering that the screening length of graphene at a heavily doped region is only a few nanometers[22]. On the other hand, the separation between the top- and bottom-layer graphene flakes is ~8 nm, and therefore the edge effect of the parallel capacitor can be excluded. The origin of the observed non-local gating effect will be discussed later in details.

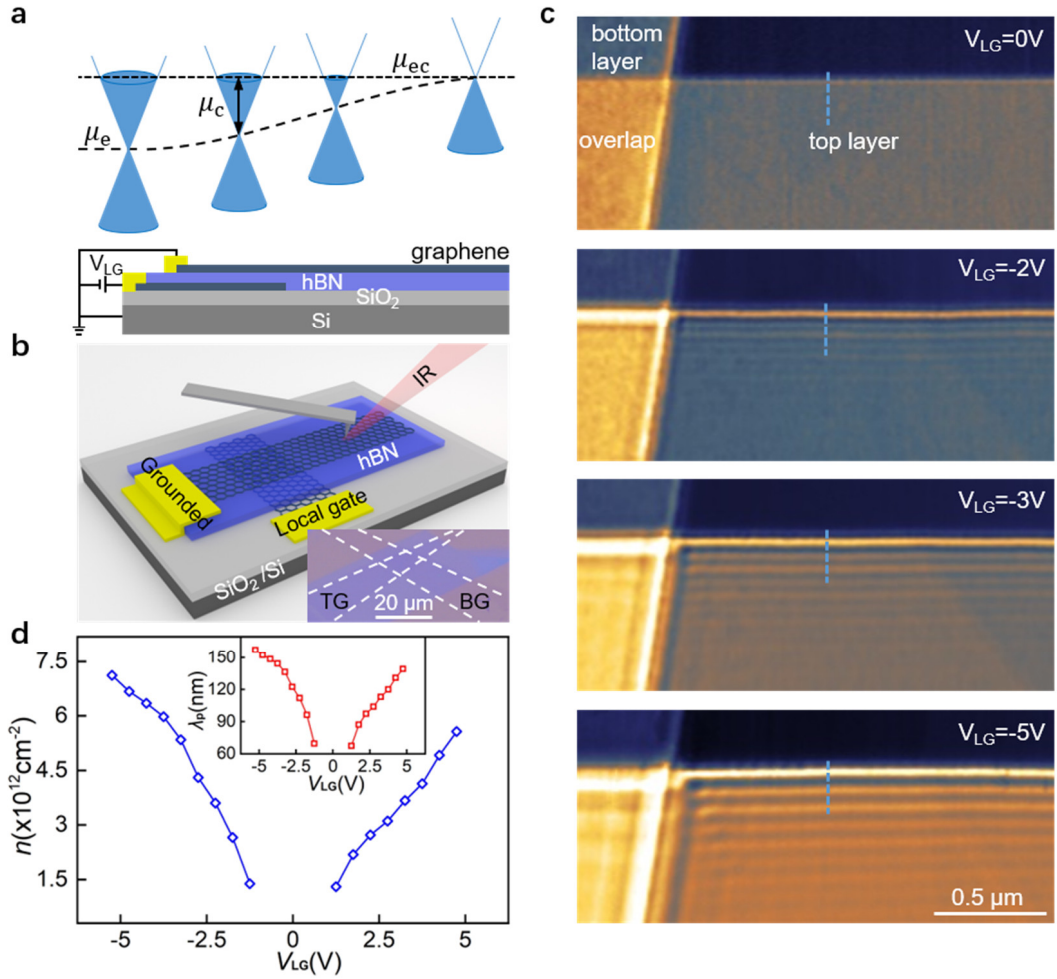

**Figure 1.** Non-local gating effect in graphene revealed by near-field infrared nano-imaging of plasmons. a) Schematic of electrical potential ($\mu_e$), chemical potential ($\mu_c$) and electrochemical potential ($\mu_{ec}$) of a partially doped graphene piece in equilibrium, where the electrochemical potential $\mu_{ec} = \mu_e + \mu_c$ keeps a constant. b) 3D diagram of the sample structure and schematic of SNOM measurement. The two crossed graphene layers (black hexagons) are separated by an insulating hBN layer with thickness of 8 nm. A laser beam of wavelength 10.6 μm was focused onto a gold-coated AFM tip (shown in gray) to excite graphene plasmons. A local-gate voltage $V_{LG}$ was applied between these two graphene layers and the near-field signal was collected at the non-overlapped graphene region. The inset shows an optical image of a typical sample. TG and BG denote the top- and bottom-layer graphene respectively and the blue region corresponds to hBN flake. c) Near-field infrared images at different local-gate voltages. The brightest region in the lower left corner corresponds to the overlapped region and the part with fringes in the right area is non-overlapped region of top-layer graphene. d) Extracted charge density versus local-gate voltage. Inset displays plasmon wavelength under different gate voltages.

To investigate the nonlocality of the gating effect, we scanned another device with very long non-overlapped graphene arm (~30 μm) as shown in **Figure 2**a. The local gate voltage is set at -5 V. Zoom-in near-field optical images at three representative areas (near the cross section, ~10 μm away, and at the far end of the graphene sheet) are shown in Figure 2b. Clear plasmon interference patterns with more than 9 fringes are observed in all three regions. It is amazing that the charge density of graphene can be tuned efficiently by a local gate placed over 30 μm away, demonstrating large nonlocality of this non-local gating effect. The plasmon profile of line-cuts in Figure 2b are plotted in Figure 2c. Plasmons wavelength is longer (~156 nm) in the region near the local gate than that further away from the local gate (~106 nm). The plasmon wavelength as a function of distance from the local gate is plotted in Figure 2d. An exponential decay of the wavelength is observed, $\lambda_p(x) = \lambda_p(0)e^{-x/x_0}$, where the attenuation length $x_0$ is about 3.3 μm, reflecting a systematic decay of the charge density and change of chemical potential in non-overlapped region with increasing the distance away from the local gate (the inset of Figure 2d).

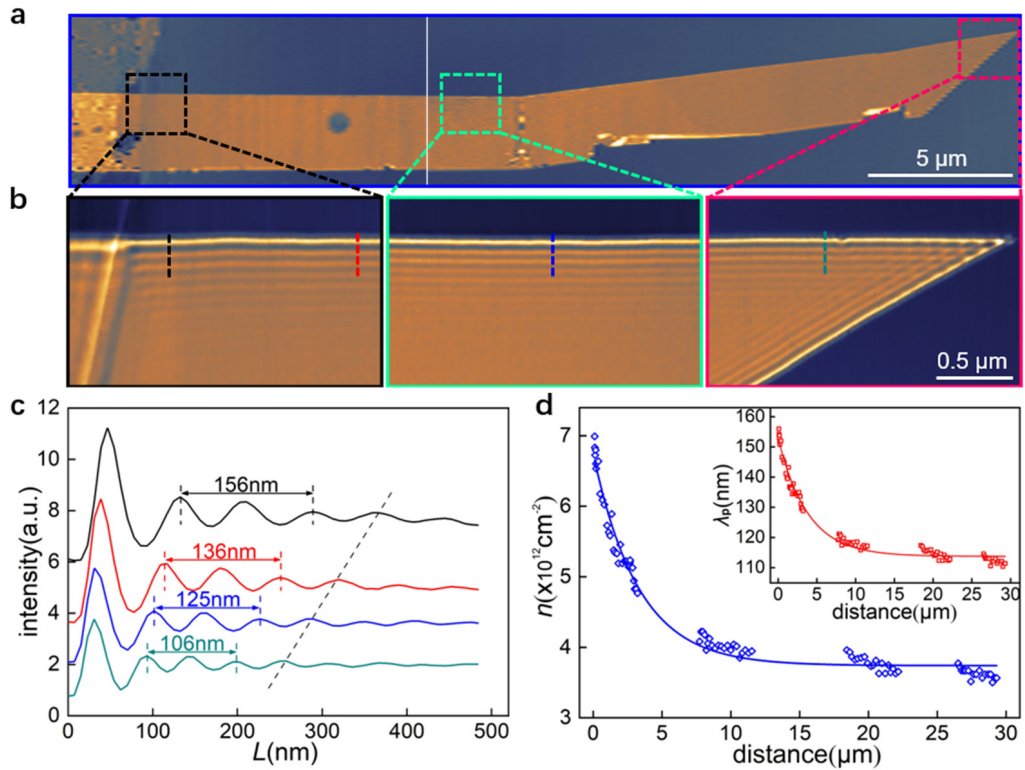

**Figure 2.** Large nonlocality of the gating effect. a) A large scale near-field optical images of top-layer graphene with ~30-μm-long non-overlapped region. b) Zoom-in images of three representative regions denoted by colored squares in a. The local gate voltage is set at -5 V. c) Line profiles of graphene plasmons at different positions denoted by colored dashed lines in b, which clearly shows a systematic change in plasmon wavelength at different positions. d) Extracted charge density at different positions of non-overlapped region. Inset is the corresponding data of the plasmon wavelength.

## 2.2 Additional evidence from KPFM and electrical transport measurements

The distribution of chemical potential ($\mu_c$) in the top-layer graphene was also detected by Kelvin probe force microscope (KPFM)[23], which is consistent with the near-field optical results. The chemical potential ($\mu_c$) of the non-overlapped region of the top-layer graphene is shown in Figure S2, where a systematic change of $\mu_c$ under different gate voltages can be clearly observed. Since $\mu_c$ is directly related to charge density, the change of $\mu_c$ reveals the tunability of the local-gate on the charge density of the non-overlapped region. The result of KPFM confirms the gate-induced non-local distribution of charge carriers.

To further confirm this peculiar non-local gating effect in a more intuitive way, electrical transport measurements were carried out[14-15]. Source and drain electrodes were fabricated on the non-overlapped top-layer graphene using standard electron beam lithography. **Figure 3**a shows the device structure and the measurement scheme. Figure 3b displays the optical image of a typical device, in which the white dashed line denotes the position of the top-layer graphene. We measured the resistance of the non-overlapped graphene region under different local-gate voltages. The transport data measured in ambient conditions (blue line in Figure 3c) shows that the non-overlapped graphene resistance can be efficiently tuned from 0.3 kΩ to 2.7 kΩ by the remote local gate. The local-gate-tunable resistance confirms unambiguously a change of charge density in the non-overlapped graphene region.

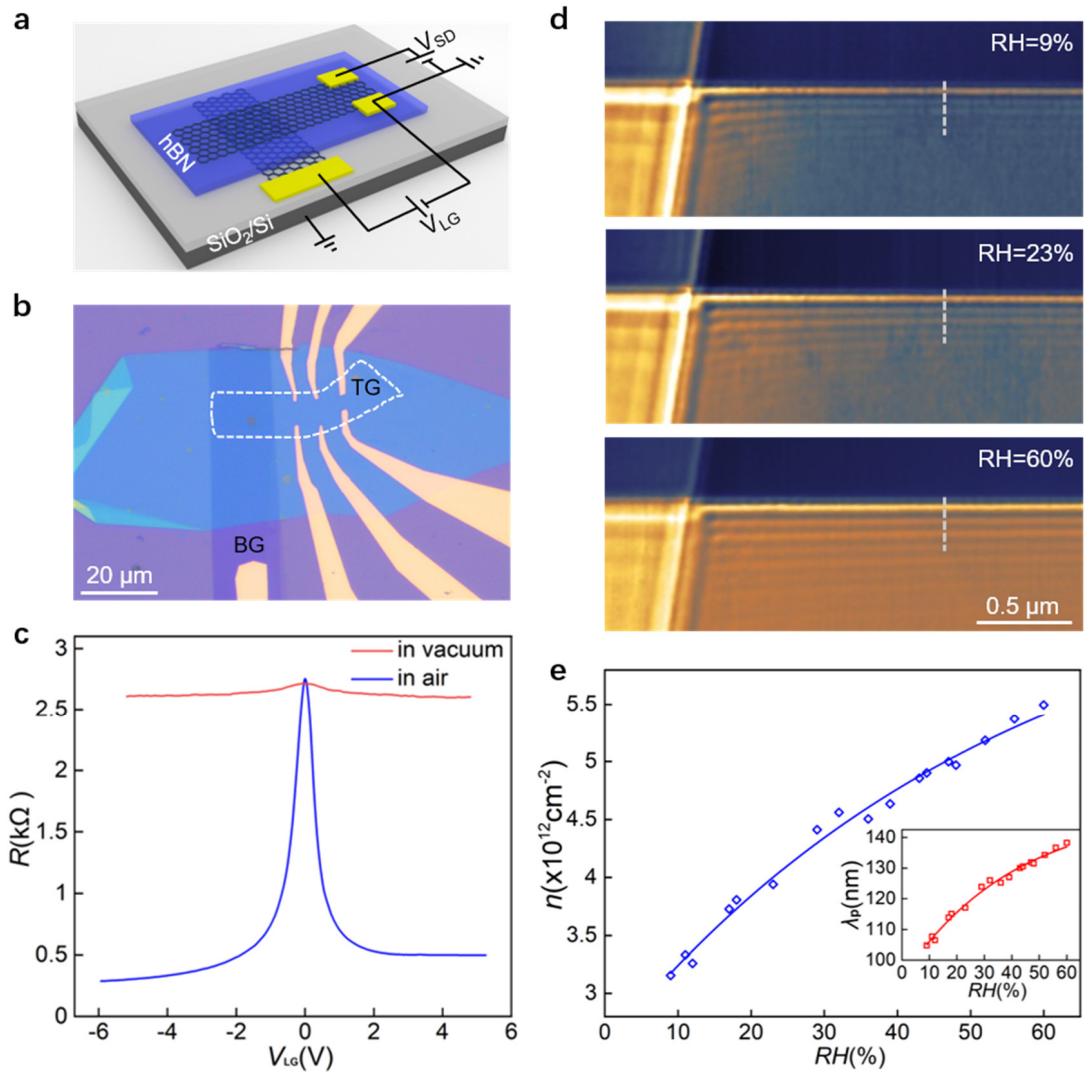

**Figure 3.** Non-local gating effect in different environments. a) Schematic of transport measurements. b) Optical image of a typical device for the transport measurements. The location of top-layer graphene is labeled by white dashed line. The insulating hBN layer is about 15 nm. c) Resistance of the non-overlapped graphene region as a function of local-gate voltage. The blue curve and red curve are recorded in air and vacuum respectively. d) Near-field optical images of plasmons at different relative humidity (RH) from 9% to 60% at a fixed local-gate voltage of -5 V. Change of plasmon wavelength and strength can be clearly seen. e) Charge density and plasmon wavelength (inset) of a specific position (gray dashed line) as a function of RH.

## 2.3 The role of water molecules

Surprisingly, the non-local gating effect disappears when the same sample was placed in vacuum. The red line in Figure 3c represents the resistance as a function of the local-gate voltage of this device measured in vacuum (~$10^{-6}$ Torr), which exhibits a constant resistance of ~2.6 kΩ. The constant resistance reflects that the charge density in the non-overlapped region is not tunable any more by the local gate in vacuum. Note that this non-local gating effect recovered completely when the same sample was placed back in ambient environment (see more details in Figure S5). This reflects that some molecules in air play an indispensable role in the non-local gating effect. It is natural to speculate that it is the water molecule that dominates the non-local gating effect. Firstly, the water molecules are polar molecules which can potentially generate polarization charges under external electric field. Secondly, the water molecules have been reported to induce hysteresis loop in low-dimensional material devices when applying electrical gating[24-26]. To verify this speculation, we purposely tune the relative humidity (RH) of the measurement environment and record plasmon wavelengths in the non-overlapped graphene region. The RH level indeed affects the plasmon wavelength as well as the non-local gating effect. Figure 3d shows near-field optical images of graphene plasmons with RH of 9%, 23% and 60% from top to bottom, respectively. The plasmon wavelengths extracted from the dashed line position in Figure 3d are plotted in Figure 3e. Evidently, the plasmon becomes stronger and its wavelength becomes longer with increasing humidity, indicating a strong positive correlation between humidity and the non-local gating effect.

Further, we directly observed the formation of water film on top-layer graphene under high humidity with local gating (See supplementary information Figure S4). Such water film indicates a large amount of absorption of water molecules on graphene surface. As polar molecules, water molecules tend to stay in a place with intense electric field in order to lower the electrical potential energy. Indeed, we found the water film was first formed near the boundary between the overlapped and non-overlapped regions where an intense in-plane electric field exists, and then extended to the far end of the non-overlapped graphene region. We also observed that the edge of the water film coincided with the edge of the region that could be tuned by the local-gate. All those phenomena show a crucial role of the water molecules in establishing the non-local gate tuning of charge density in graphene.

## 3. Discussion

Note that the remotely gated graphene region is electrically connected to the ground, so it always stays at zero electrical potential. This implies that electric field lines could not come out from the non-overlapped graphene region. Therefore, if the graphene is electrically doped by net charges, there must exist equal amount of opposite charges in the water layer that can terminate electric field lines from the charge carriers in graphene. Without such opposite charges in water layer, it is impossible to dope the grounded non-overlapped graphene region.

The above experimental observations unambiguously demonstrated that water molecules are essential for the non-local gating. In the following, we analyze possible origins of charges in the water layer and potential working mechanisms of the non-local gating effect. The charges in water layer could in principle come from ions. Ions naturally exists in water layer and can be separated by in-plane electric field built between graphene regions with different doping levels. Such ions could potentially dope the non-overlapped graphene region and minimize the electrostatic energy. However, this mechanism is not likely to be true for this case as explained below. We experimentally use ultra-pure water to generate water vapor in our vacuum chamber in the transport measurement as shown in Figure S6. We found that the non-local gating effect is still observable. At room temperature, the concentration of $H^+$ and $OH^-$ ions in pure water is typically $10^{-7}$ mol/L, which is too low to generate such a high doping level (~$10^{12}$ cm$^{-2}$) in graphene with a few monolayers of water molecules. Therefore, the ion origin of the non-local gating effect could be ruled out.

Next, we examine the possibility of polarization charges in water layer. The quantum capacitance induced in-plane electric field ($\nabla \mu_e$) in graphene is typically non-uniform. As a result, the water molecules are non-uniformly polarized ($\vec{P}_\text{water}$), and a gradient of polarization formed as shown in **Figure 4**a, which will induce polarization charges ($n_P = -\nabla \cdot P_\text{water}$) in the water layer. Considering the electrical dipole moment of a single water molecule is 1.85 Debye and assuming that the orientation of all the absorbed water molecules reverses within a transition distance of ~10 μm, one can estimate the density of polarization charges, which is in the order between $10^{10}$ cm$^{-2}$ and $10^{11}$ cm$^{-2}$. Still, the polarization charge density is kind of low and difficult to generate the experimentally observed doping level of $10^{12}$ cm$^{-2}$.

To further check this possibility more quantitatively, we have built a model to simulate the non-local gating effect. The model consists of a graphene layer, a dielectric hBN layer in the bottom, and a 1nm-thick layer of water on top, as shown in Figure 4b, in which the layer of polar molecules of water is treated as a dielectric layer that can provide polarization charges. We set the graphene with a finite charge density $n_0$ in the left terminal and charge neutral in the right end. Due to the small DOS in graphene, its chemical potential $\mu_c$ depends strongly on the charge density $n$ as[12] $\mu_c = \hbar v_F \sqrt{\pi n}$ where $\hbar$ and $e$ are the reduced Planck constant and the elementary charge, and $v_F = c/300$ is the Fermi velocity in graphene. As a result, the inhomogeneous charge density will induce a gradient in the chemical potential $\mu_c$. In equilibrium, the change of chemical potential in graphene will be compensated by an inversely changed electrical potential, as $\boldsymbol{\nabla}\mu_c = -e\boldsymbol{E}$. Combining the above two equations with the Gauss's law $\varepsilon_0 \varepsilon_r \boldsymbol{\nabla} \cdot \boldsymbol{E} = n$, we are able to simulate the local-gate induced remote-control of charge density in graphene numerically using a finite element analysis method (see methods). The simulated charge distribution is displayed in Figure 4c. The simulation (red curve) with a water layer can roughly reproduce our experimental observation. Note that the dielectric constant of water in the simulation is set to be in the order of $10^8$, which is much larger than that of typical bulk water ~$10^2$. Further theoretical studies need to be carried out to understand the largely enhanced permittivity of the ultrathin water layer on graphene. Presumably, the polarization charges from dielectric water layer may not be able to take full charge of the non-local gating effect.

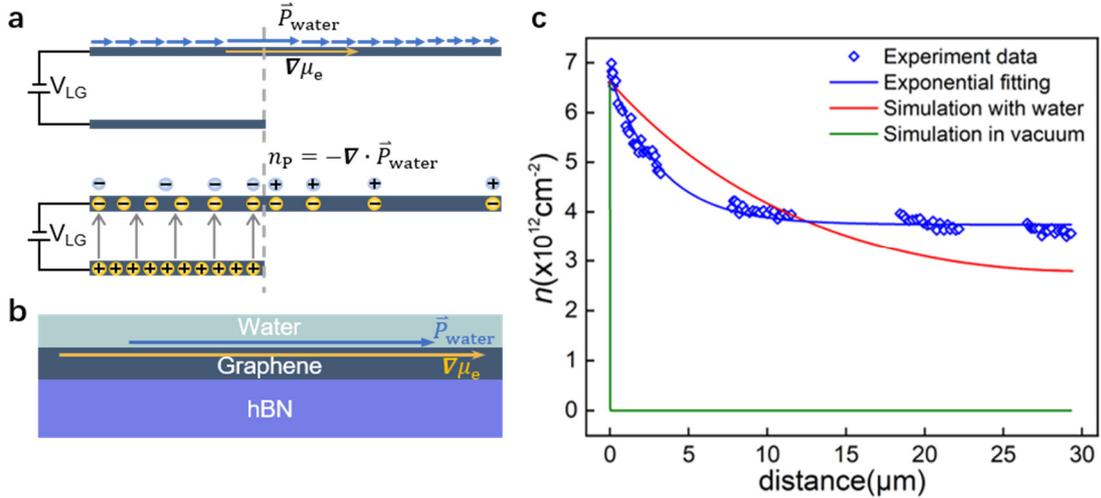

**Figure 4.** Modeling and simulation of the non-local gating effect. a) Schematic of water molecules polarization and induced polarization charges on graphene surface. $\boldsymbol{\nabla}\mu_e$ is the gradient of electrical potential, i.e. the in-plane electric field, and $n_P = -\boldsymbol{\nabla} \cdot \boldsymbol{P}_{water}$ is the polarization charges induced from the non-uniform polarization of water. b) The structure used in our simulation, consists of a hBN layer, a graphene layer and a water layer. c) Comparison between calculated and experimental results of the charge distribution in graphene with a local gate.

We then propose a complementary origin for the non-local gating effect－the charge-transfer origin. In the overlapped region, the water molecules are expected to orientate vertically to the graphene surface, under the driven force from the vertical gate electric field. When departing from the overlapped region and entering into the non-overlapped region, the water molecules would gradually alter their orientation from vertical to horizontal due to the existence of in-plane electric field in graphene as shown in **Figure 5**a. Such collective alignment and rotation of water molecules would dope graphene in two ways. Firstly, they will generate polarization charges as discussed above, which will induce opposite charges in graphene. Secondly, well-aligned water molecules can transfer net charges to graphene directly, which can provide extra doping to graphene, as described below.

It is well known that there exists charge transfer at the liquid-graphene interface[27]. We have carried out density functional theory simulations to calculate the charge transfer between a water molecule and a graphene sheet. We consider two conFigureurations: one with the two hydrogens pointing upward, and the other with the two hydrogens pointing downward to the graphene sheet, as shown in Figure 5b,c. In the first case, we observe electron-doping from water to graphene, and in the second one we observe hole-doping. Thus, bipolar doping can be induced by the absorbed water molecules. In both cases, the transferred carriers are in the $p_z$ band of the graphene, and thus can contribute to the conductivity. The amount of the charge transfer is 0.005~0.01 electron/hole per water molecule, depending on the exchange-correlation function. Considering the hydrogen-bond length is around 3 Å, we estimate that the density of water molecules at the interface is $8\times10^{14}$ cm$^{-2}$. Accordingly, the doped electron/hole density is above $10^{12}$ cm$^{-2}$, which agrees with the experimental data.

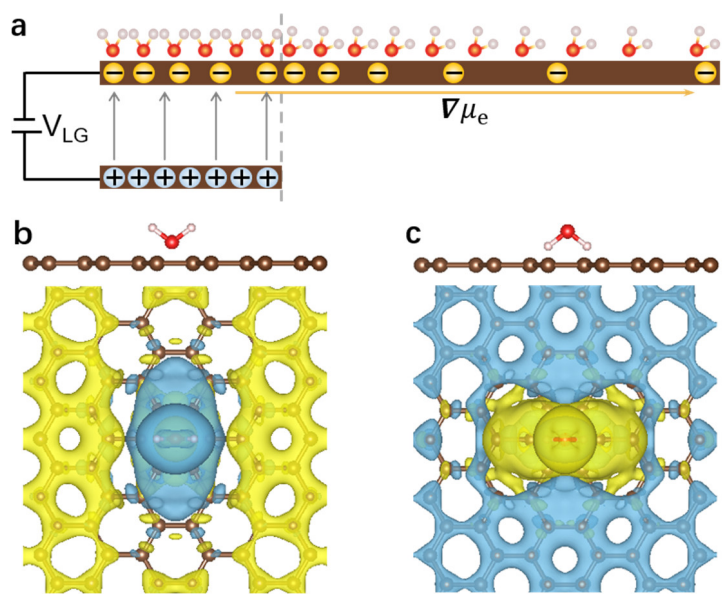

**Figure 5.** DFT calculations of the charge transfer at the water-graphene interface. a) Schematic of water molecules directional alignment on graphene surface induced by both the vertical gate electric field and the in-plane electric field. This alignment would result in the charge transfer at the water-graphene interface. b,c) Charge density differences defined as $\rho$ (water + graphene)- $\rho$ (graphene)- $\rho$ (water) for the two conFigureurations: the two hydrogens pointing upward (panel b) and the two hydrogens pointing downward (panel c). The yellow color indicates electrons and the blue color indicates holes.

We now reconsider the large effective dielectric constant $\sim 10^8$ required for producing the non-local gating phenomenon in our simulation. The large dielectric constant corresponds to a large screening effect arising from the water layer. Such an unusual large screening may be a result of the net charges transferred at the graphene-water interface. The quantum capacitance of graphene induces a finite drop in both chemical potential and electrical potential between the overlapped and non-overlapped regions. Consequently, there establishes an in-plane electric field that provides the driving force and initializes the non-local gating effect. The in-plane electric field also absorbs water molecules from environment, which can on one hand provide polarization charges and on the other hand transfer net charges to graphene. Those charges can largely screen and reduce the amplitude of the established in-plane electric field in graphene. A requirement of a super large dielectric constant of $10^8$ reflects that the in-plane electric field has been greatly reduced to be a tiny value. To keep the potential drop to a constant value under such an electrical screening, the transition area has to be expended. An enlarged transition region that extends over tens of micrometers in this experiment manifests a novel non-local gating effect.

With such a non-local gating effect, one is able to design novel functional devices with much simpler structure. For example, one can achieve electrical gating ability in a device made of suspended 2D materials, which is normally unavailable. Another example is a device for investigating plasmon propagation and reflection across a nanoscale gap between two graphene pieces with different charge densities. Two bottom-gate electrodes are required in order to independently tune the charge density of the two graphene pieces. Normally, the two gate electrodes have to be aligned perfectly with the two graphene pieces, respectively. Fabrication of such a device is extremely difficult with conventional lithographic methods due to the requirement of nanometer scale alignment. However, using the non-local gating effect, the gate electrodes do not have to be perfectly aligned with the graphene pieces. Thus, the fabrication would become much simpler.

In conclusion, we have reported experimental observations of a non-local electrostatic gating effect in graphene, in which the charge density of graphene can be effectively tuned by a remote-gate located over tens of micrometers away. The non-local gating effect is initially driven by an in-plane electric field established due to the quantum capacitance of graphene and further largely amplified by a layer of absorbed water molecules. Theoretical analysis reveals that a spatial evolution of the orientation of absorbed water molecules under external electric field could induce both polarization charges and net charge transfer to graphene, which is most likely responsible for the observed non-local gating effect. Our observations open up a window for designing novel electronic devices with remote electrical gating.

## Methods Section

**Device fabrication:** our samples were fabricated on a 500-μm-thick silicon wafer with 285 nm oxide layer. The bottom-layer graphene was exfoliated onto the wafer from bulk graphite directly. The hBN and top layer graphene were first exfoliated on Poly propylene carbonate (PPC) film and then transferred onto the bottom layer graphene to form a parallel-plate capacitor structure. After each transfer step, we treated the sample with hydrogen plasma to remove the adhesive residues. The cleaning process was conducted in a remote plasma-enhanced chemical vapor deposition (R-PECVD) system at temperature of 300 °C. The power of plasma was 30 W and the flow of $H_2$ was 35 SCCM with pressure about 47 Pa. The distance between our sample and the center of coil was about 45 cm. The gate electrodes were fabricated by depositing 3 nm Ti and 50 nm Au with a shadow mask. We scraped the sample surface with AFM tip by lift mode to remove residual contaminants before the near-field optical measurements.

**Infrared nano-imaging:** Our home-made scattering-type SNOM apparatus was based on an AFM (Bruker Innova). A beam of $CO_2$ laser with 10.6 μm wavelength was illuminated onto a gold-coated AFM tip, which generated a near-field hot spot at the nanometer-scale tip apex. Such a largely confined hot spot could provide extra momentum and excited plasmons in graphene. The excited plasmons propagated and got reflected at the graphene edges. The reflected plasmons modulated the local field at the tip apex and changed the light scattering that was recorded by a MCT detector placed in the far field. Plasmon interference pattern could be obtained when scanning over the graphene. In order to suppress background scattering from the cantilever and the sample, the tip was vibrated vertically with an amplitude of ~40 nm at a frequency of about 200 kHz, and the detector signal was demodulated at a higher order harmonic frequency by a lock-in amplifier (Zurich Instruments, HF2LI).

**COMSOL simulation:** Numerical simulations were conducted using the 2D electrostatic module of the commercial software package COMSOL. In all simulations, the electrical potential was monitored to calculate the carrier density in graphene. The boundary was the zero-charge boundary condition, indicated that $\boldsymbol{n} \cdot \boldsymbol{D} = 0$ at the boundary. The mesh size was optimized to get a convergence solution in this model. To put graphene in a 3D framework, we assumed that the thickness of graphene was 1 nm and charge density was homogeneously distributed along the out-of-plane direction in graphene so that we could transform the surface charge density $n$ to the volume charge density $n_{3D}$. The effective permittivity of graphene at zero frequency was from reference[28].

**First-Principles calculation:** Our density functional theory simulations were calculated by the Vienna *Ab-initio* Simulation Package[29]. The projector augmented wave pseudopotentials[30] were employed, and the wave functions were expanded using a plane-wave basis-set with an energy cutoff of 300 eV. Both the local density approximation[31] and the generalized gradient approximation formulated by Perdew, Burke, and Ernzerhof[32] have been employed, and the results were almost the same. The supercell size equaled to the 6×6 times of the primitive cell of the graphene. The Brillouin zone integration was sampled at the Γ-point. The van der Waals interaction between the water molecule and the graphene was described by the semi-empirical approach of Grimme[33] when using the generalized gradient functional.

## Supporting Information

Supporting Information is available from the Wiley Online Library or from the author.

## Acknowledgements

We thank F.W., W.L. and J.K. for helpful discussions. This work is supported by the National Key Research and Development Program of China (2016YFA0302001) and the National

Natural Science Foundation of China (11774224 and 11704027). Z.S. acknowledges support from the Program for Professor of Special Appointment (Eastern Scholar) at Shanghai Institutions of Higher Learning, and additional support from a Shanghai talent program. K.W. and T.T. acknowledge support from the Elemental Strategy Initiative conducted by the MEXT, Japan, Grant Number JPMXP0112101001, JSPS KAKENHI Grant Number JP20H00354 and the CREST (JPMJCR15F3), JST. We would also like to thank the Centre for Advanced Electronic Materials and Devices (AEMD) of Shanghai Jiao Tong University (SJTU) for the support in device fabrication.

**Author contributions**

Z.S. conceived this project. A.D. prepared the heterostructure samples and performed the near-field infrared measurements. A.D., P.S., J.C., B.L. fabricated the devices. A.D. performed the transport measurements. C.H. carried out the COMSOL simulations. J.M. provided the theoretical analysis and first principles calculation. K.W. and T.T. provided the hBN crystals. A.D., C.H., P.S., X.L., J.C., B.L., Q.L., J.M. and Z.S. analyzed the data. All authors discussed the results and contributed to writing the manuscript.

**Conflict interest**

The authors declare no conflict of interest.

**References**

[1] H. Yukawa, Proceedings of the Physico-Mathematical Society of Japan. 3rd Series 1935, 17, 48.

[2] A. H. Castro Neto, F. Guinea, N. M. R. Peres, K. S. Novoselov, A. K. Geim, Reviews of Modern Physics 2009, 81, 109.

[3] E. H. Hwang, S. Das Sarma, Physical Review B 2007, 75, 205418.

[4] M. B. Lundeberg, Y. Gao, R. Asgari, C. Tan, B. Van Duppen, M. Autore, P. Alonso-González, A. Woessner, K. Watanabe, T. Taniguchi, R. Hillenbrand, J. Hone, M. Polini, F. H. L. Koppens, Science 2017, 357, 187.

[5] G. Viola, T. Wenger, J. Kinaret, M. Fogelström, Physical Review B 2018, 97, 085429.

[6] L. Wang, I. Gutiérrez-Lezama, C. Barreteau, D.-K. Ki, E. Giannini, A. F. Morpurgo,

Physical Review Letters 2016, 117, 176601.

[7] J. Xia, F. Chen, J. Li, N. Tao, Nature Nanotechnology 2009, 4, 505.

[8] S. Ilani, L. A. K. Donev, M. Kindermann, P. L. McEuen, Nature Physics 2006, 2, 687.

[9] D. L. John, L. C. Castro, D. L. Pulfrey, Journal of Applied Physics 2004, 96, 5180.

[10] A. J. Bard, L. R. Faulkner, *Electrochemical methods: fundamentals and applications*, John Wiley & Sons, New York 2001.

[11] R. HillenbRand, B. Knoll, F. Keilmann, Journal of Microscopy 2001, 202, 77.

[12] Z. Fei, A. S. Rodin, G. O. Andreev, W. Bao, A. S. McLeod, M. Wagner, L. M. Zhang, Z. Zhao, M. Thiemens, G. Dominguez, M. M. Fogler, A. H. Castro Neto, C. N. Lau, F. Keilmann, D. N. Basov, Nature 2012, 487, 82.

[13] J. Chen, M. Badioli, P. Alonso-González, S. Thongrattanasiri, F. Huth, J. Osmond, M. Spasenović, A. Centeno, A. Pesquera, P. Godignon, A. Zurutuza Elorza, N. Camara, F. J. G. de Abajo, R. Hillenbrand, F. H. L. Koppens, Nature 2012, 487, 77.

[14] K. S. Novoselov, A. K. Geim, S. V. Morozov, D. Jiang, Y. Zhang, S. V. Dubonos, I. V. Grigorieva, A. A. Firsov, Science 2004, 306, 666.

[15] K. S. Novoselov, A. K. Geim, S. V. Morozov, D. Jiang, M. I. Katsnelson, I. V. Grigorieva, S. V. Dubonos, A. A. Firsov, Nature 2005, 438, 197.

[16] L. Wang, I. Meric, P. Y. Huang, Q. Gao, Y. Gao, H. Tran, T. Taniguchi, K. Watanabe, L. M. Campos, D. A. Muller, J. Guo, P. Kim, J. Hone, K. L. Shepard, C. R. Dean, Science 2013, 342, 614.

[17] F. Pizzocchero, L. Gammelgaard, B. S. Jessen, J. M. Caridad, L. Wang, J. Hone, P. Bøggild, T. J. Booth, Nature Communications 2016, 7, 11894.

[18] A. N. Grigorenko, M. Polini, K. S. Novoselov, Nature Photonics 2012, 6, 749.

[19] D. N. Basov, M. M. Fogler, F. J. García de Abajo, Science 2016, 354, aag1992.

[20] T. Low, A. Chaves, J. D. Caldwell, A. Kumar, N. X. Fang, P. Avouris, T. F. Heinz, F. Guinea, L. Martin-Moreno, F. Koppens, Nature Materials 2017, 16, 182.

[21] G. X. Ni, H. Wang, J. S. Wu, Z. Fei, M. D. Goldflam, F. Keilmann, B. Özyilmaz, A. H. Castro Neto, X. M. Xie, M. M. Fogler, D. N. Basov, Nature Materials 2015, 14, 1217.

[22] D. Wong, F. Corsetti, Y. Wang, V. W. Brar, H.-Z. Tsai, Q. Wu, R. K. Kawakami, A. Zettl, A. A. Mostofi, J. Lischner, M. F. Crommie, Physical Review B 2017, 95, 205419.

[23] W. Melitz, J. Shen, A. C. Kummel, S. Lee, Surface Science Reports 2011, 66, 1.

[24] H. Wang, Y. Wu, C. Cong, J. Shang, T. Yu, ACS Nano 2010, 4, 7221.

[25] A. Veligura, P. J. Zomer, I. J. Vera-Marun, C. Józsa, P. I. Gordiichuk, B. J. v. Wees, Journal of Applied Physics 2011, 110, 113708.

[26] P. L. Levesque, S. S. Sabri, C. M. Aguirre, J. Guillemette, M. Siaj, P. Desjardins, T. Szkopek, R. Martel, Nano Letters 2011, 11, 132.

[27] L. D'Urso, C. Satriano, G. Forte, G. Compagnini, O. Puglisi, Physical Chemistry Chemical Physics 2012, 14, 14605.

[28] E. J. G. Santos, E. Kaxiras, Nano Letters 2013, 13, 898.

[29] G. Kresse, J. Furthmüller, Physical Review B 1996, 54, 11169.

[30] G. Kresse, D. Joubert, Physical Review B 1999, 59, 1758.

[31] W. Kohn, L. J. Sham, Physical Review 1965, 140, A1133.

[32] J. P. Perdew, K. Burke, M. Ernzerhof, Physical Review Letters 1996, 77, 3865.

[33] S. Grimme, Journal of Computational Chemistry 2006, 27, 1787.